\documentclass[sigplan,9pt]{acmart}

\usepackage{amssymb}
\usepackage{newtxmath}
\usepackage{graphicx}
\usepackage[ruled,vlined]{algorithm2e}
\usepackage{geometry}
\usepackage{subfigure}
\usepackage{subcaption}
\usepackage{float}
\usepackage{subfigure}
\usepackage{svg}
\usepackage{enumitem}
\usepackage{tabularx}
\usepackage{booktabs}
\usepackage{subcaption}
\usepackage{stfloats}  

\newboolean{showcomments} 

\setboolean{showcomments}{true}
\ifthenelse{\boolean{showcomments}}
{
\newcommand{\Xinjiao}[2]{{\color{#1}Xinjiao: #2}}
\newcommand{\Ruilong}[2]{{\color{#1}Zhuoxu: #2}}
\newcommand{\Yisu}[2]{{\color{#1}Yitong: #2}}
\newcommand{\Dirk}[2]{{\color{#1}\fbox{\begin{minipage}{0.8\linewidth}Dirk: #2\end{minipage}}}}
}
{
\newcommand{\Xinjiao}[2]{}
\newcommand{\Ruilong}[2]{}
\newcommand{\Yisu}[2]{}
\newcommand{\Dirk}[2]{}
}

\setcopyright{acmlicensed}
\copyrightyear{2025}
\acmYear{2025}
\setcopyright{rightsretained}
\acmConference[APNET 2025]{9th Asia-Pacific Workshop on Networking}{August 07--08, 2025}{Shang Hai, China}
\acmBooktitle{9th Asia-Pacific Workshop on Networking (APNET 2025), August 07--08, 2025, Shang Hai, China}
\acmPrice{}
\acmDOI{10.1145/3735358.3735382}
\acmISBN{979-8-4007-1401-6/25/08}

\begin{document}

\title{Rethinking Dynamic Networks and Heterogeneous Computing with Automatic Parallelization}

\author{Ruilong WU, Xinjiao Li, Yisu Wang, Xinyu Chen, Dirk Kutscher}
\affiliation{%
  \institution{The Hong Kong University of Science and Technology (Guangzhou)}
  \city{Guangzhou}
  \state{Guangdong}
  \country{China}
}

\email{{rwu408, xli886, ywang418}@connect.hkust-gz.edu.cn, {xinyuchen, dku}@hkust-gz.edu.cn}



\begin{abstract}
Hybrid parallelism techniques are essential for efficiently training large language models (LLMs). Nevertheless, current automatic parallel planning frameworks often overlook the simultaneous consideration of node heterogeneity and dynamic network topology changes, limiting their effectiveness in practical applications. In this paper, we address these limitations by modeling heterogeneous nodes within dynamically changing network environments and leveraging simulation-based strategies to determine optimal parallel configurations. Our approach enables fine-grained workload allocation tailored for heterogeneous nodes and complex network scenarios, achieving performance competitive with state-of-the-art methods under regular and stable network conditions. Additionally, we introduce a strategy pruning technique to rapidly discard infeasible parallel configurations, substantially reducing the search space and accelerating the search process through parallel execution within the simulator. Preliminary evaluations confirm that our method notably enhances training performance on heterogeneous nodes and demonstrates improved adaptability in complex, dynamic scenarios such as cloud computing environments.
\end{abstract}
\vspace{-20pt}


\vspace{-40pt}
\vspace{-40pt}
\vspace{-40pt}
\vspace{-40pt}
\vspace{-40pt}


\maketitle

\section{Introduction}

The rapid growth in parameter count of deep neural networks (DNNs), especially large language models (LLMs)\cite{radford2019language}\cite{brown2020language}\cite{touvron2023llama}\cite{ouyang2022training}
\cite{guo2025deepseek}\cite{liu2024deepseek} based on the Transformer\cite{vaswani2017attention} architecture, has made distributed parallel training across large GPU clusters indispensable. Thus, efficient implementation of distributed training is critical. Researchers have proposed various parallel strategies\cite{narayanan2021efficient}
\cite{rajbhandari2020zero}\cite{rasley2020deepspeed}, including Data Parallelism (DP)\cite{li2020pytorch}, Tensor Parallelism (TP)\cite{narayanan2021efficient}, Pipeline Parallelism (PP)\cite{huang2019gpipe}\cite{fan2021dapple}
\cite{li2021chimera}\cite{liu2023hanayo}\cite{feng2023mobius}, Sequence Parallelism (SP)\cite{korthikanti2023reducing}, and Fully Sharded Data Parallelism (FSDP)\cite{zhao2023pytorch}, to address computational, storage, and communication challenges in training large models. However, selecting appropriate parallel strategies in practical large-scale clusters typically requires extensive manual tuning. While existing automatic search frameworks, such as ALPA\cite{zheng2022alpa}, AMP\cite{li2022amp}, Metis\cite{um2024metis} and Galvatron\cite{miao2022galvatron}, offer some degree of automation, their deployment in real-world scenarios is limited due to overly idealized assumptions.

This paper aims to address the problem of selecting distributed parallel strategies more effectively for realistic scenarios. Our core insight is that \textbf{computation can be viewed as mapping data and algorithms onto computational devices, while communication corresponds to data transmission tasks across network links}. Specifically, by selecting suitable parallel strategies, computational tasks can be efficiently assigned to heterogeneous computing devices and communicated through network links. However, due to device performance heterogeneity and dynamic network conditions, the actual execution time of tasks typically exhibits significant uncertainty.

The uncertainty in task execution time arises mainly from two aspects: first, variability in computation and communication times caused by heterogeneous device performance and fluctuations in network bandwidth; second, additional variations resulting from operators' splitting and fusion processes. For example, operator fusion reduces memory accesses and thus shortens execution time, while operator splitting can effectively utilize idle computational resources, also reducing execution time. Additionally, decomposing traditional collective communication operations such as \texttt{all-reduce} into \texttt{reduce-scatter} and \texttt{all-gather} can significantly enhance communication efficiency.

To address these practical challenges, we propose an integrated optimization framework combining strategic operator splitting and fusion, an adaptive task scheduling strategy based on parallelized branch-and-bound search, and resource management strategies tailored for heterogeneous computational environments and dynamic network conditions. We validate our framework using SimAI\cite{wangsimai}, an existing performance prediction model, and demonstrate significant performance improvements over mainstream frameworks. Specifically, our contributions include:

\vspace{-4pt}
\begin{itemize}
\item A novel multi-edge physical link abstraction model that more accurately describes heterogeneous device connectivity characteristics and link contention conditions;
\item A parallelized branch-and-bound optimization algorithm that systematically searches task scheduling strategies, significantly improving task execution efficiency;
\item Preliminary experimental validation using SimAI, indicating the potential of our method to outperform existing mainstream frameworks under heterogeneous computational environments and dynamic network conditions.
\end{itemize}
\vspace{-8pt}

\section{Background and Motivation}

\begin{figure}[htbp]
    \centering
    \includegraphics[width=1\linewidth]{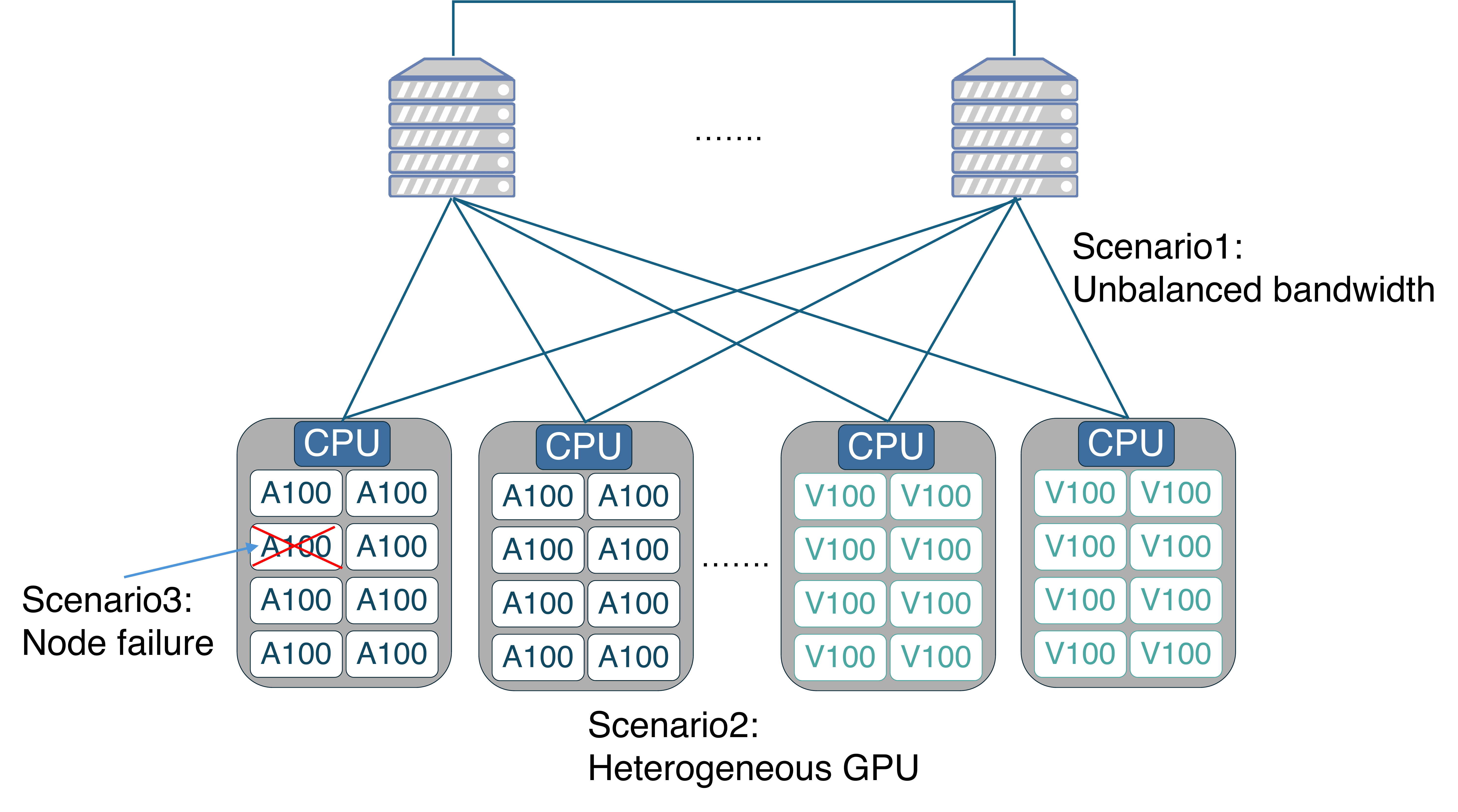}
    \caption{Representative Scenarios on hybrid parallelism: S1 Dynamic Bandwidth Variations; S2 Heterogeneous GPU Performance; S3 Device changes and connection adjustments caused by node failures}
    \label{moti}
\end{figure}

This section addresses critical challenges faced in realistic distributed training environments, specifically illustrated by the scenarios depicted in Figure \ref{moti}. In practical GPU clusters, several factors significantly impact overall training efficiency and robustness: (1) heterogeneous GPU setups combining diverse device types, (2) unbalanced network bandwidth causing performance bottlenecks, and (3) node failures resulting in computational disruptions. 

In Section 2.1, we first examine the impact of GPU performance heterogeneity on overall system throughput and discuss predictive performance modeling approaches. In Section 2.2, we analyze dynamic network conditions, emphasizing the necessity for adaptive bandwidth management and fault-tolerance mechanisms. Lastly, in Section 2.3, we explore strategic operator fusion and splitting methods, highlighting their potential to effectively mitigate performance degradation and improve resource utilization under these challenging conditions.

\vspace{-12pt}
\subsection{Performance Heterogeneity}

Performance heterogeneity refers to variations in computational speed and capabilities among devices of the same type. Even if all nodes within a cluster employ GPUs that share the same instruction set (e.g., CUDA), significant performance disparities can still exist due to differences in microarchitecture or hardware generation.

\begin{figure}[htbp]
    \centering
    \includegraphics[width=1\linewidth]{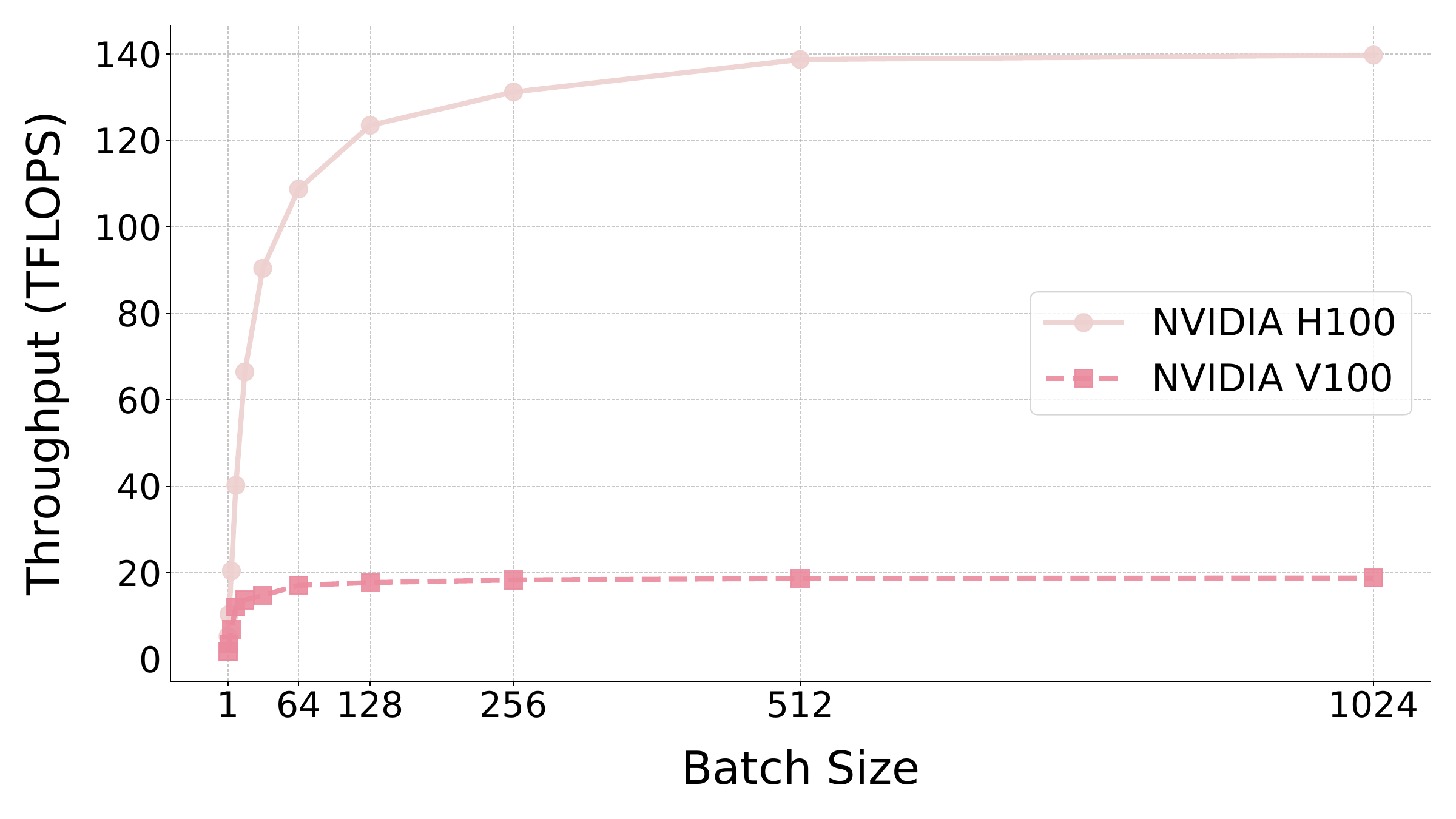}
    \caption{Attention Throughput: H100 vs V100}
    \label{10}
\end{figure}

The Roofline Model\cite{10.1145/1498765.1498785} is commonly used to analyze and predict computational system performance. It characterizes the performance of a system using the following equation:

\begin{equation}
\text{roofline}_{BW} = \min(K \times \text{memBW}_p, \text{FLOPs}_p)
\end{equation}

where $\text{FLOPs}_p$ is the peak floating-point operations per second and $\text{memBW}_p$ is the peak memory bandwidth of the GPU. The term $K$ represents the arithmetic intensity, defined as the number of floating-point operations per memory access, computed as:

\begin{equation}
K = \frac{\text{FLOPs}_k}{\text{mem}_k}
\end{equation}

Figure \ref{10} illustrates throughput differences between H100 and V100 GPUs executing the same attention kernel. We observe significant computational capability differences between these GPUs. Once the computational load reaches a certain threshold, the GPU throughput stabilizes at a constant value.

However, the Roofline Model has limitations in accurately modeling fused operators and operations with explicitly specified resource usage, as actual GPU execution times are heavily influenced by specific hardware attributes. Prior research\cite{273786} \cite{10.1145/3669940.3707265} has employed Multi-Layer Perceptron (MLP) to estimate GPU performance, addressing performance as a \textbf{nonlinear}, \textbf{multivariate} function. In such scenarios, traditional optimization methods like Integer Linear Programming (ILP) and Dynamic Programming (DP) struggle to effectively map variable-length operators onto devices. This limitation arises fundamentally because ILP and DP cannot solve optimization problems with nonlinear objectives in convex spaces.

In contrast, simulators precisely predict CUDA kernel execution times, providing accurate operator execution time estimates. Moreover, simulators can concurrently evaluate execution times for multiple scheduling strategies, significantly accelerating the identification of optimal parallelization strategies.

\subsection{Dynamic Networks}
Dynamic networks are characterized by topological changes over time, contrasting with static networks that maintain constant nodes and edges. Formally, a dynamic network can be modeled as a temporal graph, represented by a sequence G(0), G(1), \dots, G(t), or a time-dependent edge set E(t).

Efficient parallel training of large language models (LLMs) across multiple GPUs inherently faces dynamic network conditions. Communication bandwidth fluctuates due to hardware limitations or network congestion, while long-running tasks frequently experience node slowdowns or failures.

\vspace{-2pt}
\subsubsection{Dynamic Bandwidth Variations}

In practical distributed training scenarios, the available bandwidth among nodes and within nodes frequently fluctuates rather than remaining constant. Such variations stem from multi-tenant data center networks, hardware bottlenecks, and background workloads. However, current distributed training frameworks\cite{li2020pytorch}\cite{abadi2016tensorflow}
\cite{shoeybi2019megatron}\cite{jiang2020unified} typically cause GPUs with higher bandwidth lanes to idle, waiting for GPUs with lower bandwidth lanes to complete data transmission, despite similar computational capabilities.

\vspace{-2pt}
\subsubsection{Dynamic Node and Interconnect Adjustments}

In long-running, large-scale training tasks, node failures or temporary disconnections are inevitable. Traditional approaches typically halt training upon encountering node failures, reloading from checkpoints, and restarting new nodes, resulting in substantial downtime and wastage of computational resources. Recent research has emphasized fault-tolerance capabilities, which enable distributed training systems to operate continuously despite node additions or removals.
For instance, ReCycle\cite{gandhi2024recycle} leverages the redundancy inherent in data-parallel training by dynamically reallocating workloads from failed nodes to the remaining active nodes, avoiding delays from node replacement. Oobleck\cite{jang2023oobleck} proactively computes pipeline-parallel configurations optimized for varying numbers of nodes, seamlessly transitioning to smaller-scale configurations upon node removal, thereby eliminating the need for retraining.

\subsection{Operation Fusion and Split}

Modern machine learning frameworks \cite{li2020pytorch}\cite{zhao2023pytorch}\cite{222575}\cite{10.1145/3341301.3359630} typically accelerate computation by forming efficient fused kernels through the fusion of multiple consecutive operators, thereby reducing data movements from external memory. A classic example is \textit{Flash Attention}\cite{dao2022flashattentionfastmemoryefficientexact}, which combines originally independent operations such as \texttt{matmul}, \texttt{dropout}, \texttt{softmax}, and \texttt{mask} into a single fused kernel, significantly shortening execution time.

\begin{figure}[htbp]
    \centering
    \includegraphics[width=1\linewidth]{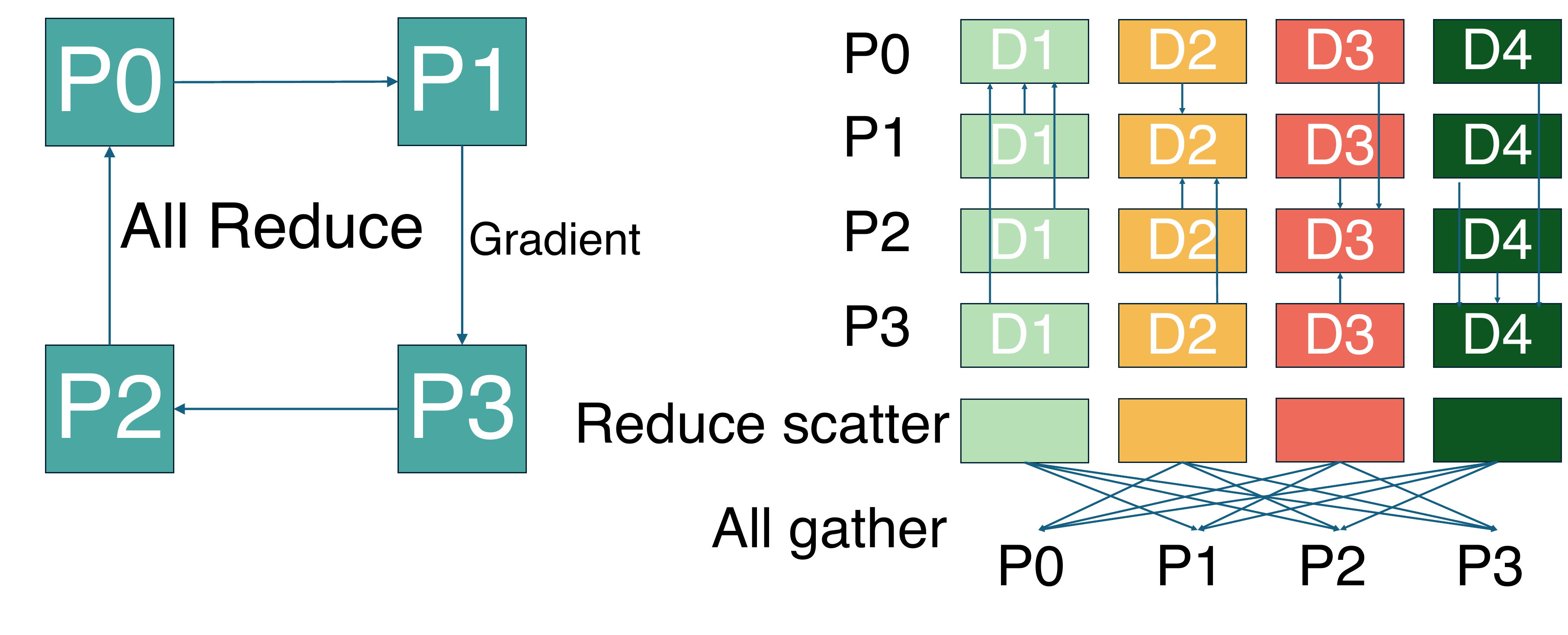}
    \caption{Comparison between traditional All-Reduce and decomposed All-Reduce (Reduce-Scatter followed by All-Gather).}
    \label{all-reduce}
\end{figure}

In contrast to operator fusion, distributed computations often utilize an operator-splitting strategy, exemplified by decomposing the standard \texttt{All-Reduce} operation into two sub-operations: \texttt{Reduce-Scatter} and \texttt{All-Gather}. As illustrated in Figure~\ref{all-reduce}, the traditional \texttt{All-Reduce} aggregates gradients fully at a single node before broadcasting the result to all other nodes. The decomposed approach, however, first partitions and aggregates gradients across nodes via the \texttt{Reduce-Scatter} step, and subsequently disseminates these partial aggregation results to all nodes through the \texttt{All-Gather} step. This decomposition effectively eliminates single-node bottlenecks and enhances overall communication efficiency.

Moreover, the process of splitting and recombining operators introduces additional opportunities for optimization in parallel computation. By decomposing operators into smaller sub-operations and recombining them in novel configurations, new operators with varying execution characteristics emerge. Searching through these configurations enables identification of optimal mappings onto heterogeneous hardware, thus effectively mitigating the previously discussed straggler effect, and ultimately leading to more balanced workload distribution and improved overall performance.

\begin{figure*}[htbp]
    \centering
    \includegraphics[width=1\textwidth]{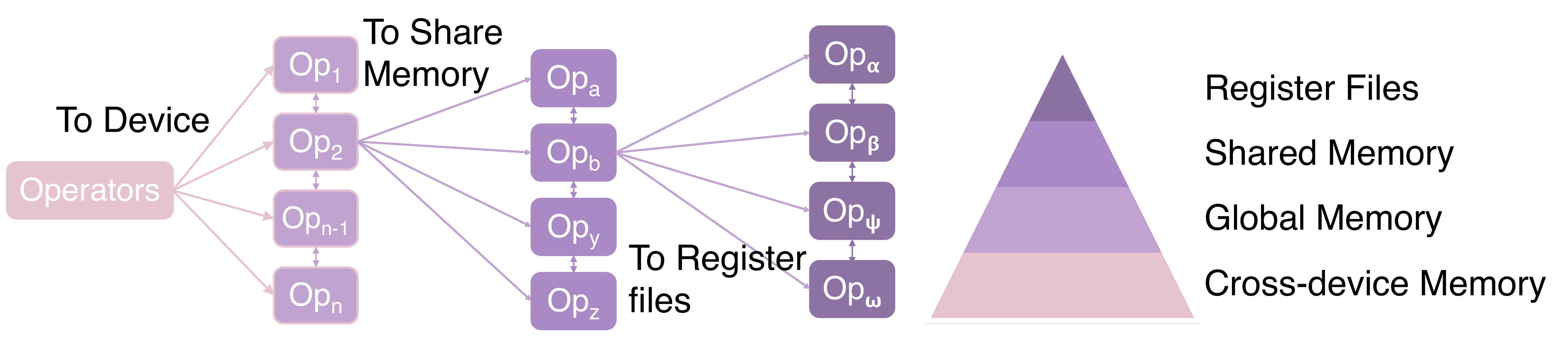}
    \caption{Hierarchy of GPU memory bandwidth optimization levels. The first level represents inter-device connections, providing a bandwidth ranging from several GB/s to tens of GB/s. The second level indicates global memory, with a bandwidth typically ranging from hundreds of GB/s to approximately 1 TB/s. The third level corresponds to the shared memory, which offers a bandwidth of several TB/s. Finally, the highest level is the register file, delivering the greatest bandwidth, typically tens of TB/s.}
    \label{design}
\end{figure*}

\section{Design}
We extend the traditional optimization approach of tensor programs\cite{10.1145/3341301.3359630}\cite{wu2024miragemultilevelsuperoptimizertensor}\cite{Hu_2024} to automatic parallel strategies across devices. Our core idea is to split operators into lower-level sub-operators and then recombine them into new operators using a simulator to predict their execution times. Constraints such as data dependencies, memory size, and bandwidth are considered to identify the optimal parallel strategy. Unlike previous methods that focused purely on parallel strategy search or heterogeneous computing, our method bridges the gap between tensor-program optimization and model parallelism. As illustrated in Figure \ref{design}, GPU devices possess four distinct memory hierarchy levels, each of which offers different bandwidth characteristics. The placement of data at these varying levels directly influences the computational efficiency. Given the heterogeneous environment and dynamic network conditions, operator execution times vary across different devices and their interconnections, and this variation does not adhere to simple linear relationships. Consequently, our approach—splitting operators first and then fusing them—allows the discovery of superior strategies. Although our method could theoretically support deeper hierarchical optimizations, our current work focuses only on first-level optimization, specifically, splitting models across different devices and searching at the global memory level to obtain optimal parallel strategies.

In Section 3.1, we introduce a Multi-Edges Assumption, and in Section 3.2, we introduce the Problem Formulation. In Section 3.3, the proposed algorithm is discussed. In Section 3.4, we discuss the search space used in our design.

\vspace{-10pt}
\subsection{Multi-Edges Assumption}

\begin{figure}[htbp]
  \centering
  \subfigure[Unequal bandwidth in DGX H100]{
    \includegraphics[width=0.62\linewidth]{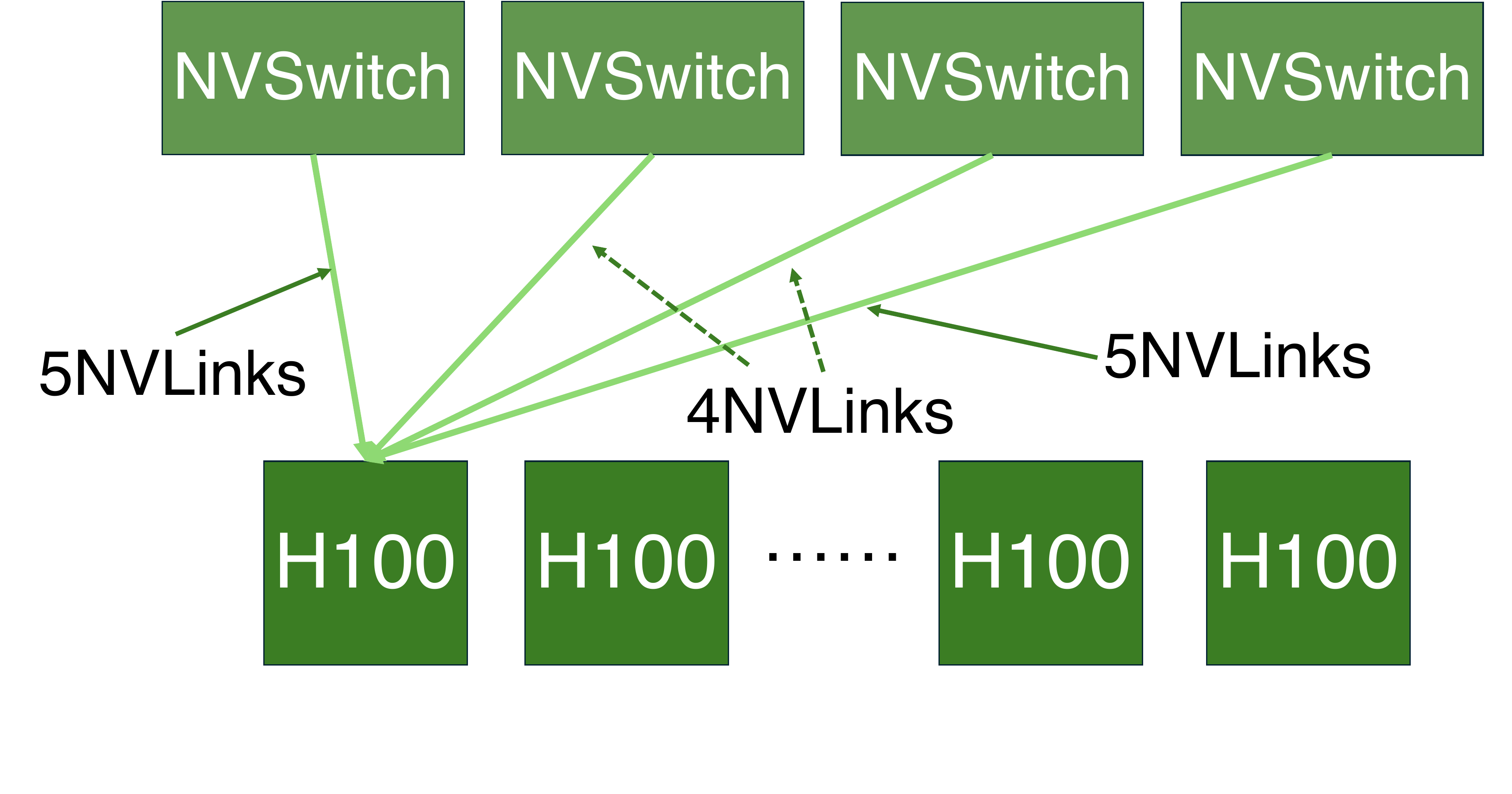}
    \label{H100}
  }
  \hfill 
  \subfigure[Conflicting connections between C2C NVLink and PCIe]{
    \includegraphics[width=0.33\linewidth]{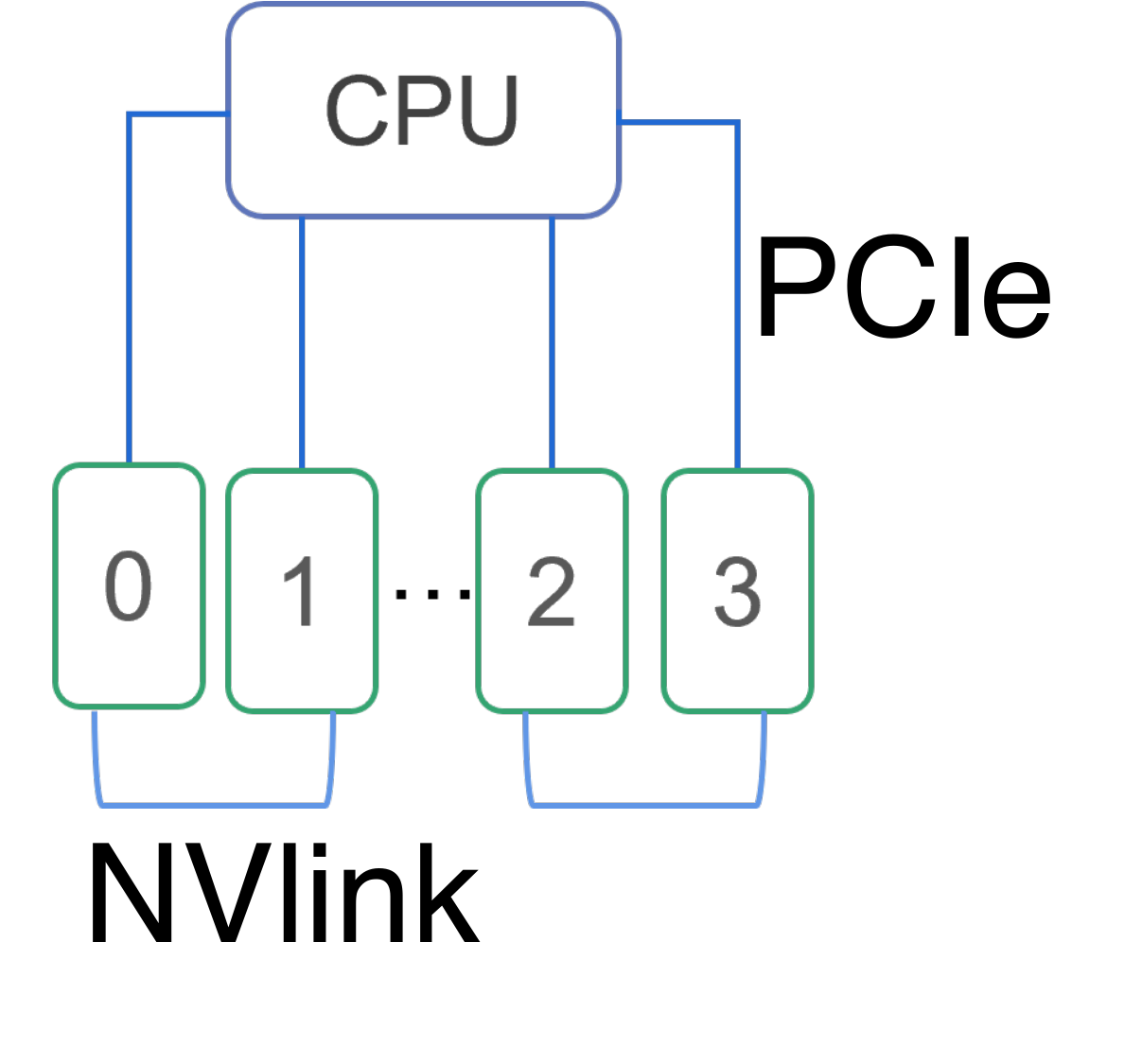}
    \label{C2C}
  }
  \caption{Two typical types of links: (a) unequal bandwidth, (b) conflicting connections.}
  \label{fig:3}
\end{figure}

The introduction of multi-edges is motivated by the fact that in real-world scenarios, a single device often has multiple physical links to other devices. As illustrated in Figure \ref{H100}, within a DGX H100\cite{NVIDIA2023DGXH100} system, the connections from each GPU to the NVSwitch are uneven, and the NVSwitches located on both sides have more ports and higher bandwidth connections, indicating that modeling these interconnected paths as equivalent connections may lead to significant discrepancies in transmission time. In addition, in NVIDIA DGX servers\cite{NVIDIA2023DGXH100}\cite{NVIDIA2017DGX1V100}\cite{NVIDIA2020Ampere}\cite{NVIDIA2017DGX1System}, the NVSwitch can perform simple arithmetic operations, which may reduce transmission bandwidth and thus influence transmission times. Similarly, Google’s TPU\cite{jouppi2023tpu}\cite{9351692} employs a torus/mesh architecture that provides multiple interconnected paths across different dimensions. Moreover, as shown in Figure \ref{C2C}, typical NVIDIA GPUs offer NVLink C2C and PCIe connections simultaneously. Although the \texttt{cudaMemcpy} function defaults to using the NVLink connection unless explicitly disabled by invoking \texttt{cudaDeviceDisable PeerAccess}, for NVIDIA GPUs, the NVLink and PCIe transfers cannot be simultaneously activated within the same kernel execution. Given these considerations, introducing a multi-edge design to explicitly model multiple physical links as potentially concurrent or conflicting network resources is essential. This approach accurately simulates parallel transmissions, avoids overly simplistic single-bandwidth assumptions, and properly manages link contention states during data-transfer scheduling.

\vspace{-2pt}
\subsection{Problem Formulation}

We formulate an operator-level scheduling and resource allocation problem for distributed DNN tasks across a heterogeneous multi-edge device graph. The goal is to optimally schedule computational operators considering their data dependencies and recombination possibilities to minimize the total weighted completion time for training multiple models. The execution and communication times are deterministically predicted using a simulation-based performance model.

\subsubsection{Input Specification}

Formally, we represent the computational graph for each model $i$ as $G_C^i = (V_C^i, E_C^i)$, where $V_C^i$ denotes atomic computational operators (e.g., convolution, matrix multiplication), and $E_C^i$ specifies data dependencies (defining execution order among operators).

The set of available heterogeneous computing devices is denoted by $V_D = { d_1, \dots, d_M }$, comprising GPUs, TPUs, and similar hardware units with diverse computational capacities and memory resources. Devices are interconnected via multi-edge physical links, represented as $L_{(d_j,d_k)}$, each with multiple bandwidth capacities $B_{(d_j,d_k)}^\alpha$ reflecting concurrent communication channels with varying bandwidth and latency.

Operator execution time $T_{\text{exec}}(v, d_j)$ for operator $v$ on device $d_j$ is obtained from a simulation model based on device-specific characteristics. The communication duration

$T_{\text{comm}}(\text{size}, \ell_{\alpha})$ indicates the time required to transfer data of a given size through link $\ell_{\alpha}$.

Memory constraints include operator execution memory, $\mathrm{Mem}{op}(v)$, representing memory usage during operator execution, and intermediate data memory, $\mathrm{Mem}{data}(u,v)$, required for storing data transferred from operator $u$ to $v$.

\subsubsection{Output Specification}

The solution comprises:
\begin{itemize}
\item Device assignment $D(v)$ for each operator $v$, with execution start $s_v$ and end $e_v$ times.
\item Selected communication link $\ell_{(u,v)}$ and the respective start $s_{(u,v)}$ and end $e_{(u,v)}$ times for each data dependency $(u,v)$.
\end{itemize}

\subsubsection{Optimization Objective}

The objective is to minimize the weighted sum of the completion times (makespans) for all models:
\begin{equation}
\label{eq:objective}
\min_{D,s,\ell} \sum_{i=1}^{n} \omega_i \cdot T_{\text{compl}}^i,
\quad \text{where } T_{\text{compl}}^i = \max_{v \in V_C^i} e_v.
\end{equation}

Here, $\omega_i$ denotes the priority weight for model $i$, and each model’s completion time is determined by its final operator’s end time.

\subsubsection{Constraints}

The primary constraints include:
\begin{itemize}
\item \textbf{Data Dependency Constraints:} An operator $v$ can begin only after all predecessors and their data transfers are completed:
\begin{equation}
\label{eq:data_dependency}
s_v \ge \max{ e_{u_j}, e_{(u_j,v)} }, \quad \forall u_j \in \text{predecessors}(v).
\end{equation}

\item \textbf{Communication Constraints:} Data transfer for dependency $(u,v)$ commences only after operator $u$ completes execution:
\begin{equation}
\label{eq:communication}
s_{(u,v)} \ge e_u.
\end{equation}

\item \textbf{Memory Constraints:} Memory usage on device $d_j$ must not exceed its total capacity $M_{d_j}$:
\begin{equation}
\label{eq:memory}
\sum_{v \in V_{C,j}} \mathrm{Mem}_{op}(v) + \sum_{(v,w) \in E_{C,j}} \mathrm{Mem}_{data}(v,w) \le M_{d_j}.
\end{equation}

\item \textbf{Bandwidth Constraints:} Total bandwidth usage on each link $\ell_{\alpha}$ at time $t$ must not exceed its bandwidth limit $B_{\alpha}$:
\begin{equation}
\label{eq:bandwidth}
\sum_{c \in C(t, \alpha)} \mathrm{rate}(c) \le B_{\alpha}.
\end{equation}

\end{itemize}

\subsection{Algorithm}
Algorithm~\ref{algo} presents our parallel branch-and-bound search method to efficiently explore the optimal operator assignment and scheduling across heterogeneous computing resources.

Initially, the algorithm starts with an initialization procedure, where input data, including computation graphs, device specifications, and resource constraints, are processed. A root node, representing an initial state with all operators unassigned, is created (line 2). The best solution and its upper bound (minimal known cost) are initialized, optionally leveraging a heuristic greedy strategy to quickly provide a baseline solution (line 4). A priority queue is then established to organize exploration nodes by their estimated cost (line 5).

The main parallel exploration procedure (\texttt{ParallelSearch}, lines 6–16) proceeds by iteratively examining nodes from the priority queue. At each iteration, the node with the minimal estimated completion cost is selected (line 8). If this node represents a complete assignment (i.e., all operators are assigned), the algorithm compares its cost against the current best-known solution, updating the latter if an improvement is found (lines 9–10).

If the node is incomplete, the algorithm generates feasible child nodes representing possible next assignments of operators, considering current scheduling constraints (lines 12–13). Each feasible child node undergoes a cost estimation procedure (line 13). Nodes whose estimated cost exceeds the current best-known upper bound are discarded early, reducing unnecessary exploration (lines 14–15).

The process repeats until all promising solutions have been explored or pruned, ultimately returning the optimal scheduling solution (line 16).

\begin{algorithm}[t]
\caption{Parallel Branch-and-Bound Search}
\label{algo}
\SetKwFunction{Init}{Initialize}
\SetKwFunction{Search}{ParallelSearch}
\SetKwData{bestUB}{best\_UB}
\SetKwData{bestSol}{best\_solution}
\SetKwData{PQ}{PQ}
\SetKwData{F}{F}
\SetKwProg{Fn}{Function}{:}{}

\Fn{\Init{}}{
    Read input graphs, devices, constraints\;
    Create root node $N_{\text{root}}$ (all operators unassigned)\;
    $\bestUB \leftarrow +\infty$, $\bestSol \leftarrow \emptyset$\;
    (Optional) Greedy initialization for $\bestSol$, $\bestUB$\;
    $\PQ \leftarrow \text{PriorityQueue()}$, $\PQ.push(N_{\text{root}})$\;
}

\Fn{\Search{$\PQ$}}{
    \While{$\PQ \neq \emptyset$}{
        $N \leftarrow \PQ.pop()$\;
        \If{$N$ is complete solution and $F(N) < \bestUB$}{
            $\bestSol \leftarrow N$, $\bestUB \leftarrow F(N)$\;
            \textbf{continue}\;
        }
        \For{each feasible child $N_{\text{child}}$ of $N$}{
            Estimate cost $F(N_{\text{child}})$\;
            \If{$F(N_{\text{child}}) < \bestUB$}{
                $\PQ.push(N_{\text{child}})$\;
            }
        }
    }
    \Return $\bestSol$\;
}
\end{algorithm}

\vspace{-2pt}
\subsection{Search Space}
Our method considers multiple splitting strategies at the operator level, where each operator has a maximum of $K$ possible splits, resulting in a combinational complexity of $\mathcal{O}(K^{|V_C^i|})$. At the device level, each resulting sub-operator needs to be mapped onto one of the available devices in $|V_D|$, adding another layer of complexity, up to $\mathcal{O}(|V_D|^{p})$, where $p$ is the number of sub-operators. Furthermore, scheduling and communication sequences must be arranged temporally, which causes the overall search space to grow exponentially.

To cope with this immense search space, we first apply constraints to eliminate infeasible choices. Subsequently, heuristic rules are utilized to effectively reduce the initial search scope, such as presetting initial search points based on multi-edge graph structures and GPU performance. Additionally, techniques such as multi-threading can be employed to concurrently simulate and evaluate multiple scenarios, significantly accelerating the search process.

\begin{figure*}[htbp]
    \centering
    \subfigure[RTX4090D and L20 GPUs]{
        \includegraphics[width=0.31\textwidth]{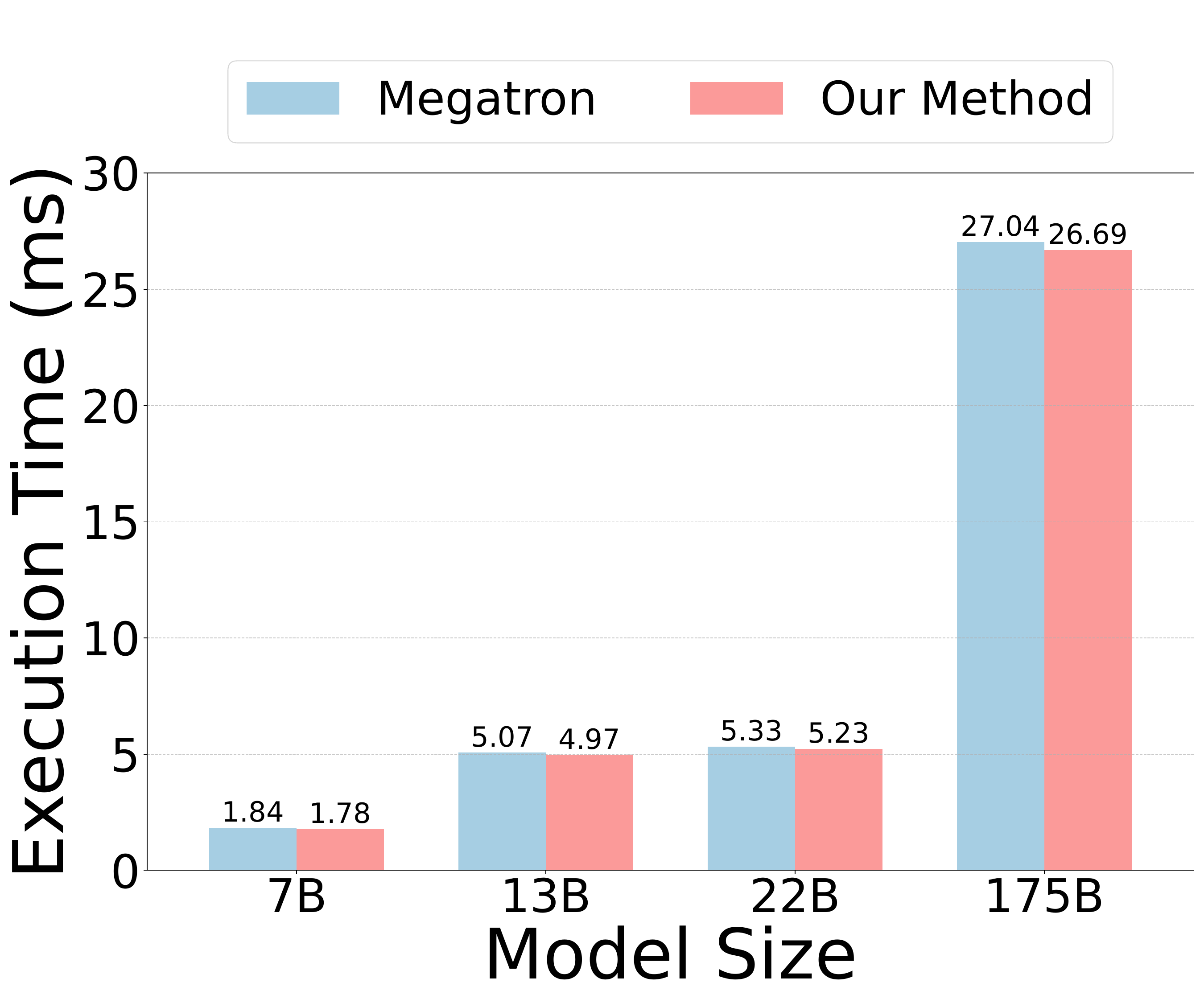}
        \label{a}
    }
    \hfill
    \subfigure[RTX4090D and V100 GPUs]{
        \includegraphics[width=0.31\textwidth]{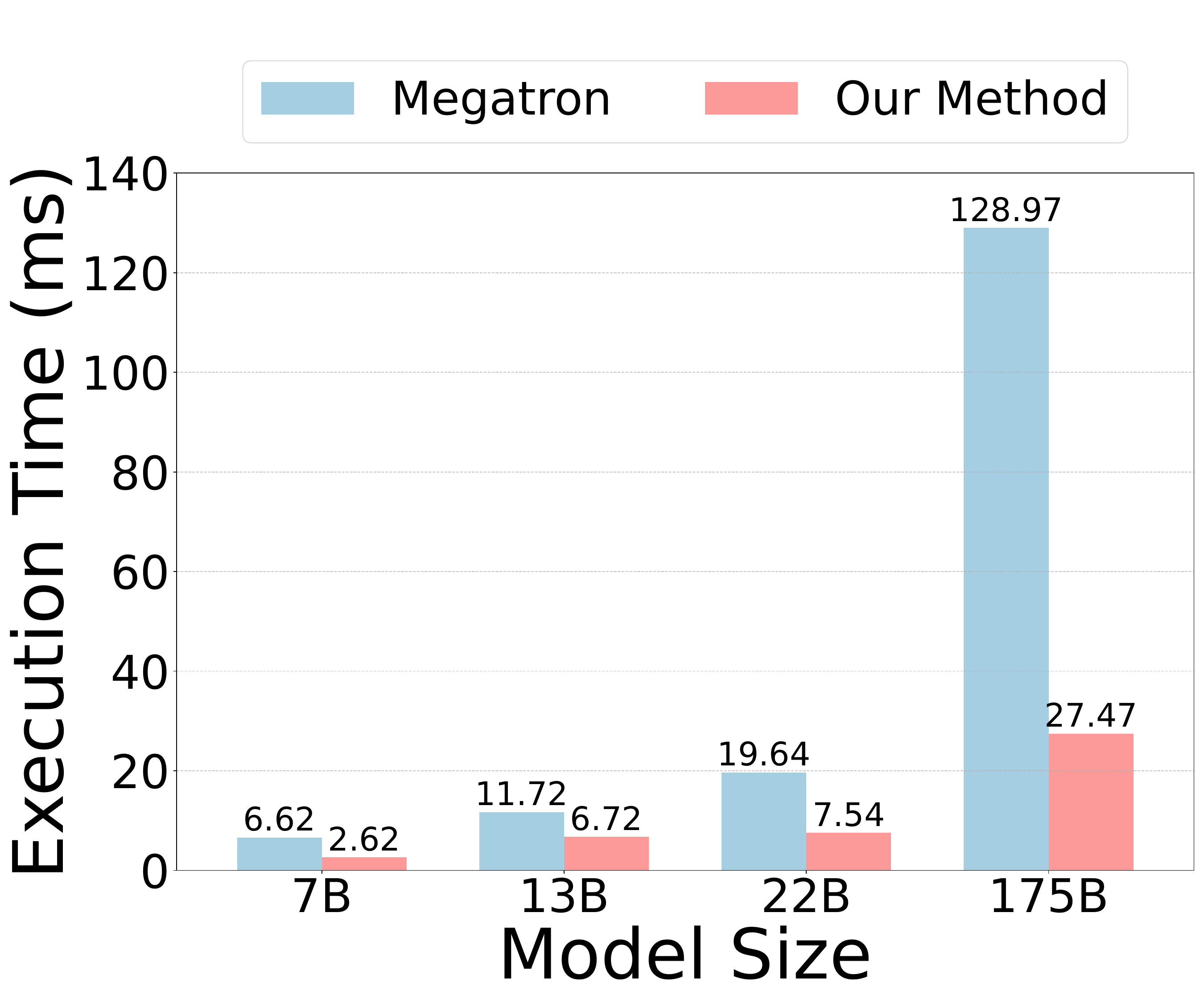}
        \label{b}
    }
    \hfill
    \subfigure[Impact of network bandwidth and parallel strategies]{
        \includegraphics[width=0.31\textwidth]{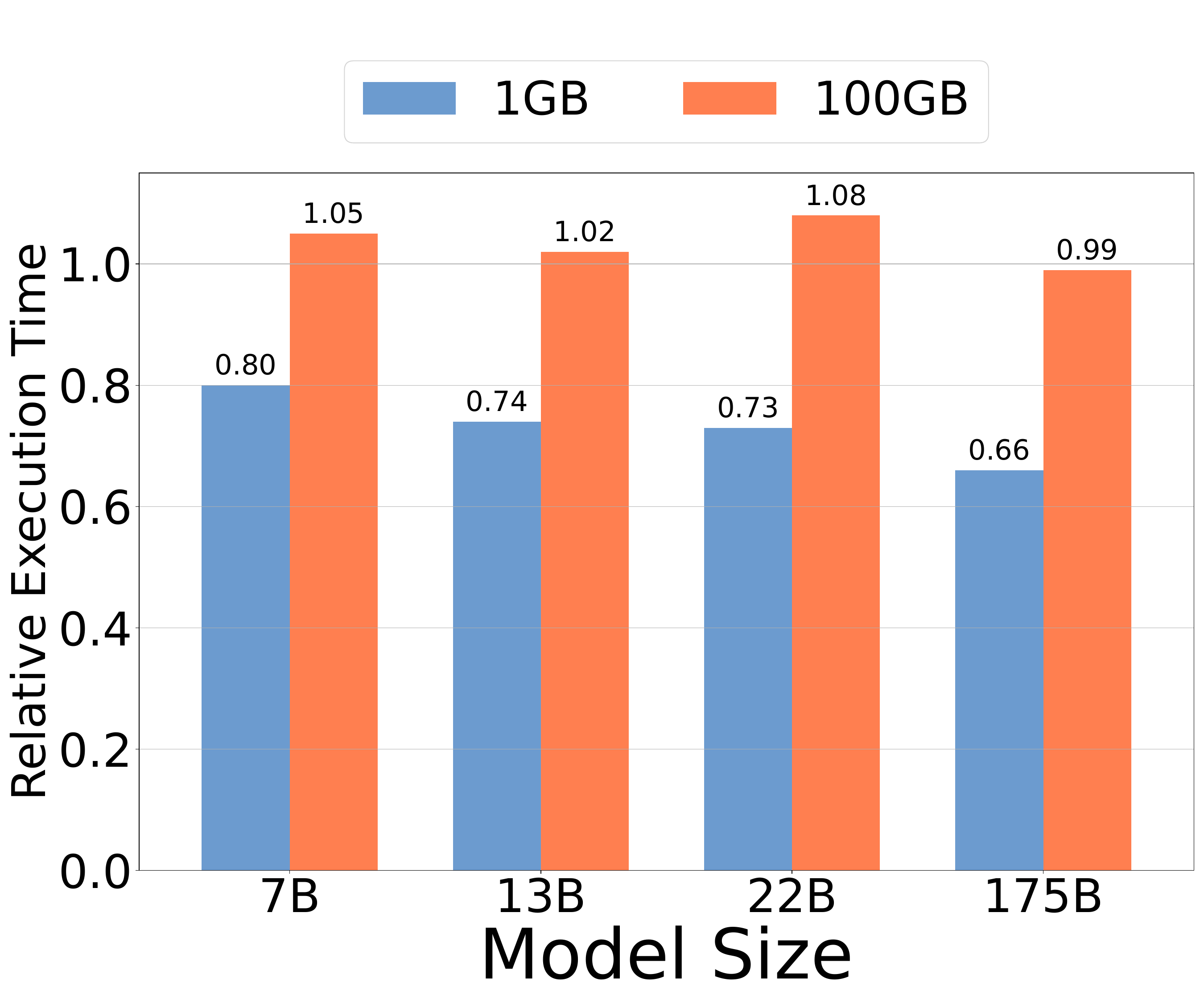}
        \label{c}
    }
    \caption{
        Comparison of execution times for training one epoch under heterogeneous computing devices and dynamic network conditions:
        (a) RTX4090D combined with L20 GPUs,
        (b) RTX4090D combined with V100 GPUs, and
        (c) relative execution times for different tensor parallelism (TP) sizes under varying network bandwidth conditions.
    }
    \label{eval}
\end{figure*}
\vspace{-8pt}
\section{Evaluation}

We use SimAI to evaluate the effectiveness of our proposed approach. The evaluation considers two scenarios: heterogeneous computation and dynamic network conditions.

\textbf{Environment Setup.}
Specifically, we utilize SimAI to simulate task execution times by leveraging the simulated execution results to guide the parameter optimization. SimAI's AIOB invokes CUDA kernel simulations for three distinct real GPUs. Subsequently, the AICB module generates workloads for further simulations and analyses. The GPUs used in this study are as follows:
\begin{itemize}[left=0pt]
\item \textbf{RTX4090D}: Features 14,592 CUDA cores based on the Ada Lovelace architecture, 114 Streaming Multiprocessors (SMs), a boost frequency of 2.52 GHz, 24 GB GDDR6X memory, and fabricated using TSMC's 5nm process.
\item \textbf{L20}: Equipped with 11,776 CUDA cores based on the Ada Lovelace architecture, 92 SMs, a boost frequency of 2.52 GHz, 48 GB GDDR6 memory, and also utilizes TSMC's 5nm process.
\item \textbf{V100}: Contains 5,120 CUDA cores based on the Volta architecture, 80 SMs, a boost frequency of 1.38 GHz, 32 GB HBM memory, and is manufactured using TSMC's 12nm process.
\end{itemize}

\textbf{Models and Baselines.} 
To demonstrate that our approach effectively identifies efficient plans, we evaluate four representative LLMs: \texttt{LLaMA\_7B}, \texttt{GPT\_13B}, \texttt{GPT\_22B}, and \texttt{GPT\_175B}. We compare our method with \texttt{Megatron} using its default configuration as a baseline.
We consider two evaluation scenarios: heterogeneous computation and dynamic network conditions.

\subsection{Scenario 1: Heterogeneous Computation.}
We evaluate the performance on clusters comprising 8, 16, 32, and 256 nodes (each containing an equal number of two GPU types) under conditions of both similar and significantly different device performance.
Although the RTX4090D and L20 GPUs originate from the same wafer, differences in the enabled functional modules result in minor performance variations. As shown in Figure \ref{a}, compared with conventional task allocation with equal computational workloads, our approach achieves approximately 1.01 to 1.03 times better performance than the general-purpose Megatron framework by utilizing layer-level task assignment.

For environments with substantial performance disparities, such as integrating the latest RTX4090D GPUs with older V100 GPUs, significant differences in the computation times exist. Consequently, our method substantially enhances performance, achieving speedups ranging from approximately 1.74 to 4.69 times compared with Megatron, as illustrated in Figure \ref{b}.

\vspace{-8pt}
\subsection{Scenario 2: Dynamic Network Conditions.}

We evaluate the execution time for training one epoch across different parallel strategies under varying network conditions using 8, 16, 64, and 256 V100-32G-PCIe GPUs. Figure \ref{c} shows the absolute execution time under different network conditions when comparing lower TP sizes to higher TP sizes. Specifically, we set TP sizes of 2 versus 4 for the 7B model, 4 versus 8 for the 13B model, 8 versus 16 for the 22B model, and 16 versus 32 for the 175B model.

We observe that, for models with fewer parameters, a lower network bandwidth increases the execution time by approximately 25\% to 52\%. This indicates that communication overhead significantly slows down training under low-bandwidth conditions, with higher TP sizes causing even greater increases in the execution time. Conversely, when the network bandwidth is not constrained, larger TP sizes still slightly increase the execution time by approximately 2\% to 8\%. However, for very large models, the overhead introduced by higher TP parallelism is offset by the non-overlapping communication time inherent in pipeline parallelism (PP).
\vspace{-8pt}

\section{Limitation}

Our study has the following two limitations:

First, due to simulator constraints, we are currently limited to parallel strategies defined by the Megatron-LM. In future work, we plan to implement a native operator-level simulator rather than relying on coarse-grained model-level task assignments.

Second, to accommodate variable-length operators, the search space grows exponentially. Even with multithreading, it is difficult for the CPU to match the search efficiency of existing methods. We plan to leverage FPGA-based acceleration for simulations in future work.

\section{Conclusion}
This study proposes a multi-edge abstraction to address task execution for LLMs under realistic dynamic network conditions and heterogeneous computing environments. Through experiments, we demonstrate that operator splitting effectively reduces task execution times in heterogeneous settings. Our method significantly outperforms existing frameworks, achieving notable speedups of up to 4.69 times. Moreover, the network conditions significantly influence the search results for parallel strategies. Under varying network environments, newly identified parallel strategies can reduce execution times by up to 52\% compared with previously optimal strategies.

\begin{acks}
This work was supported by the Guangdong Provincial Project (2023QN10X048, 2023ZT10X009), the Guangzhou Municipal Key Laboratory on Future Networked Systems (2024A03J0623), the Guangzhou Municipal Science and Technology Project (2023A03J0011), and the Natural Science Foundation of China (U23A20339). We thank the support from PulseBeam. Xinyu Chen and Dirk Kutscher are corresponding authors.
\end{acks}
\bibliographystyle{ACM-Reference-Format}
\newpage
\bibliography{sample-base}


\end{document}